\documentclass[prl,twocolumn,showpacs]{revtex4}
\usepackage{graphicx}
\usepackage{graphics}
\usepackage{amsfonts}
\usepackage{amsmath}
\usepackage{epsfig}

\begin{document}
\catcode`\ä = \active \catcode`\ö = \active \catcode`\ü = \active
\catcode`\Ä = \active \catcode`\Ö = \active \catcode`\Ü = \active
\catcode`\ß = \active \catcode`\é = \active \catcode`\è = \active
\catcode`\ë = \active \catcode`\ô = \active \catcode`\ê = \active
\catcode`\ø = \active \catcode`\ò = \active \catcode`\í = \active
\catcode`\Ó = \active \catcode`\ú = \active \catcode`\á = \active
\catcode`\ã = \active
\defä{\"a} \defö{\"o} \defü{\"u} \defÄ{\"A} \defÖ{\"O} \defÜ{\"U} \defß{\ss} \defé{\'{e}}
\defè{\`{e}} \defë{\"{e}} \defô{\^{o}} \defê{\^{e}} \defø{\o} \defò{\`{o}} \defí{\'{i}}
\defÓ{\'{O}} \defú{\'{u}} \defá{\'{a}} \defã{\~{a}}



\newcommand{\li}{$^6$Li}
\newcommand{\na}{$^{23}$Na}
\newcommand{\vect}[1]{\mathbf #1}
\newcommand{\g}{g^{(2)}}
\newcommand{\one}{|1\rangle}
\newcommand{\two}{|2\rangle}
\newcommand{\V}{V_{12}}

\title{Observation of Bose-Einstein Condensation of Molecules}

\author{M.W. Zwierlein, C.A. Stan, C.H. Schunck, S.M.F. Raupach, S. Gupta, Z. Hadzibabic, and W. Ketterle}

\affiliation{Department of Physics\mbox{,} MIT-Harvard Center for Ultracold Atoms\mbox{,}
and Research Laboratory of Electronics,\\
MIT, Cambridge, MA 02139}
\date{November 26, 2003; revised December 5, 2003}

\begin{abstract}
We have observed Bose-Einstein condensation of molecules.  When a
spin mixture of fermionic \li \ atoms was evaporatively cooled in an
optical dipole trap near a Feshbach resonance, the atomic gas was
converted into \li$_2$ molecules.  Below 600 nK, a Bose-Einstein
condensate of up to 900,000  molecules was identified by the sudden
onset of a bimodal density distribution. This condensate realizes the
limit of tightly bound fermion pairs in the crossover between BCS
superfluidity and Bose-Einstein condensation.
\end{abstract}

\maketitle

Over the last few years, many different approaches have been used to
cool and trap molecules \cite{molecules,donl02mol}. One major goal
has been the creation of molecular Bose-Einstein condensates which
could lead to advances in molecular spectroscopy, studies of
collisions, and precision tests of fundamental symmetries.

Recently, a new technique for creating ultracold molecules led to
major advances towards molecular Bose-Einstein condensation.
Molecules were produced from ultracold atoms
\cite{xu03na_mol,durr03mol,herb03cs_mol,rega03mol,cubi03,stre03,joch03lith}
near a Fesh\-bach resonance \cite{feshbach}, where a molecular state
is resonant with the atomic state and molecules can form without heat
release. These molecules are highly vibrationally excited and would
usually undergo fast decay. However, in the case of fermionic atoms
the molecules showed very long lifetimes
\cite{cubi03,stre03,joch03lith,rega03lifetime}.  This has been
attributed to Pauli suppression of the vibrational quenching process
which couples a very weakly bound molecular state to much more
tightly bound lower lying vibrational states \cite{petr03dimers}. We
have now been able to cool such molecules to Bose-Einstein
condensation.

This Bose-Einstein condensate represents one extreme of the crossover
from Bose-Einstein condensation of tightly bound pairs (molecules) to
BCS superfluidity of Cooper pairs, where fermions form delocalized
pairs in momentum space \cite{becbcs}.

In most of the recent experiments, molecules were formed by sweeping
an external magnetic field through the Feshbach resonance,
adiabatically converting atoms to molecules
\cite{rega03mol,stre03,herb03cs_mol,durr03mol,xu03na_mol}. This
atom-molecule coupling is a coherent two-body process
\cite{coherent}.

In the case of $^6$Li, experimental work indicated
\cite{cubi03,joch03lith}, and theoretical work predicted
\cite{kokk03atom_mol,chin03therm} that cooling the atoms at constant
magnetic field would create an atom-molecule mixture in thermal
equilibrium. In this case, the atoms and molecules are coupled by
three-body recombination \cite{threebody}. For temperatures lower
than the binding energy of the molecular state, an almost pure
molecular gas should form, and at even lower temperatures, a
molecular Bose-Einstein condensate. This work demonstrates that this
surprisingly simple method to create molecular condensates works. The
success of this approach depends on a very favorable ratio of
collisional rates for formation and decay of molecules which may be
unique to $^6$Li.

The goal of molecular BEC was reached in several steps.  Using
Feshbach resonances, atomic condensates were put into an
atom-molecule superposition state \cite{donl02mol}. Pure molecular
gases made of bosonic atoms were created close to \cite{herb03cs_mol}
or clearly in \cite{xu03na_mol} the quantum-degenerate regime, but
the effective heating time (of about 2 ms in Ref.~\cite{xu03na_mol})
was too short to reach equilibrium. Earlier this month, while this
work was in progress, two papers were submitted.
Ref.~\cite{grei03mol_bec} observed a quantum degenerate gas of
potassium molecules with an effective lifetime of 5 to 10 ms,
sufficiently long to reach equilibrium in two dimensions and to form
a non-equilibrium or quasi condensate \cite{shva02non_eq_bec}.
Ref.~\cite{joch03mol_cold} provided indirect evidence for a
long-lived condensate of lithium molecules \cite{footnote2}. Here we
observe the formation of a condensate by evaporative cooling of a
molecular gas close to equilibrium.

The basic scheme of the experiment is similar to our earlier work
when we identified two Feshbach resonances in lithium by monitoring
the loss of trapped atoms due to three-body recombination as a
function of the external magnetic field \cite{diec02fesh}. This
process leads to ultracold molecules in the highest vibrational state
below the continuum. However, no attempt was made to detect these
molecules because until very recently \cite{cubi03,joch03lith} they
were believed to decay rapidly.

Our experimental setup was described in
Refs.~\cite{diec02fesh,hadz03big_fermi}. After laser cooling and
sympathetic cooling with sodium atoms in a magnetic trap, 35 million
lithium atoms in the $|F,m_F\rangle =|3/2,3/2\rangle$ state were
transferred into an optical trap formed by a single far detuned laser
beam with up to 7 W of power at 1064 nm. The beam had a 20 $\mu$m
waist and was aligned horizontally along the symmetry axis of the
magnetic trap. This generated a 650 $\mu$K deep trapping potential
with 15 kHz radial and 175 Hz axial trapping frequencies. They were
determined with an accuracy of 10 \% by exciting dipolar oscillations
with an atomic sodium condensate and scaling them to lithium atoms using the
ratios of polarizabilities and masses.

The $^6$Li atoms were then transferred to the lowest energy state
$|1\rangle$, using an adiabatic frequency sweep around the lithium
hyperfine splitting of $228\,$MHz. DC magnetic fields of up to
$1025\,$G could be applied, a range encompassing the
$|1\rangle-|2\rangle$ Feshbach resonance
\cite{diec02fesh,bour03,gupt03rf} where $|2\rangle$ denotes the
second lowest hyperfine state of $^6$Li.

Most of our experiments were performed at a magnetic field of
$770\,$G.  This was below, but still within the width of the broad
Feshbach resonance. Here the atomic scattering length is positive
corresponding to a stable weakly bound molecular state.  Using
RF-induced transitions near $80\,$MHz, an equal mixture of states
$|1\rangle$ and $|2\rangle$ was prepared with a ratio of temperature
$T$ to Fermi temperature $T_F$ around 0.3.  The sample was cooled in
350~ms by ramping down the laser power of the optical trap to
typically $10^{-3}$ of the maximum power resulting in a calculated trap depth for unbound atoms of
650 nK.  The weakly bound molecules have twice the atomic
polarizability. They experience the same trap frequencies and twice
the trap depth as the lithium atoms. Therefore, we expect mainly
atoms to be evaporated.

\begin{figure}
    \begin{center}
    \includegraphics[width=3.2 in]{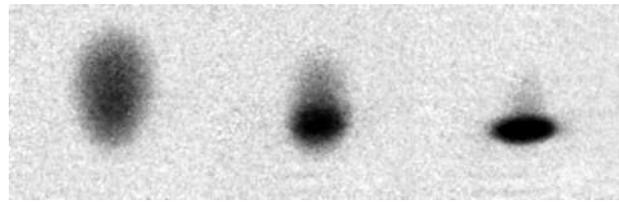}
    \caption[Title]{Observation of Bose-Einstein condensation in a
    molecular gas.  Shown are three single-shot absorption images after 6 ms of
    ballistic expansion for progressively lower temperatures (left to right).  The
    appearance of a dark spot marks the onset of BEC.
    The field of view
    for each image is 1.4 mm x 1.4 mm.  The long axis of the optical dipole
    trap was vertical in the image.
    \label{fig:2D_images}}
    \end{center}
\end{figure}

\begin{figure}
    \begin{center}
    \includegraphics[width=2.5 in]{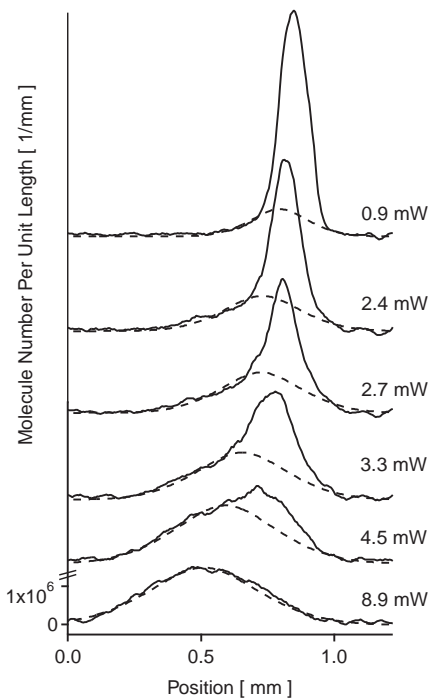}
    \caption[Title]{Bimodality in the density distribution of a molecular gas.
    Shown are radially averaged profiles through absorption images like those in
     Fig.~\ref{fig:2D_images}, as a function of final laser power. The dashed lines are fits to the thermal clouds.
    \label{fig:waterfall}}
    \end{center}
\end{figure}

Atoms and molecules were detected by absorption imaging after
ballistic expansion times of 1 to 30 ms. During the time-of-flight,
the magnetic field was suddenly switched off, and atoms in both
states were imaged simultaneously since the two optical transition frequencies
are equal at zero field. Molecules were detected by first
dissociating them by sweeping the magnetic field  across the Feshbach
resonance up to 925~G, and then by imaging the resulting atoms at
zero field. With the Feshbach sweep, molecules and residual atoms
were imaged together.  Without it, only the unbound atoms were
detected after switching off the magnetic field.  We have found that
during the initial phase of the evaporative cooling the atomic
population dominated.  A significant molecule fraction formed around
$T \sim 2 \mu$K, and in the final phase of the cooling, no atoms
could be discerned. The absorption images and profiles shown in
Figs.~\ref{fig:2D_images} and \ref{fig:waterfall} therefore represent
purely molecular column densities.

When the laser power of the optical dipole trap was ramped down, the
shrinking size of the cloud in absorption imaging signaled lower
temperatures. Very abruptly, the smooth distribution changed to a
bimodal distribution --- the well-known ``smoking gun'' of
Bose-Einstein condensation \cite{ande95,davi95bec}
(Figs.~\ref{fig:2D_images} and \ref{fig:waterfall}).  Due to a slight
asymmetry of our trapping potential, the centers of the condensate
and of the thermal cloud were shifted.

\begin{figure}
    \begin{center}
    \includegraphics[width=2.7 in]{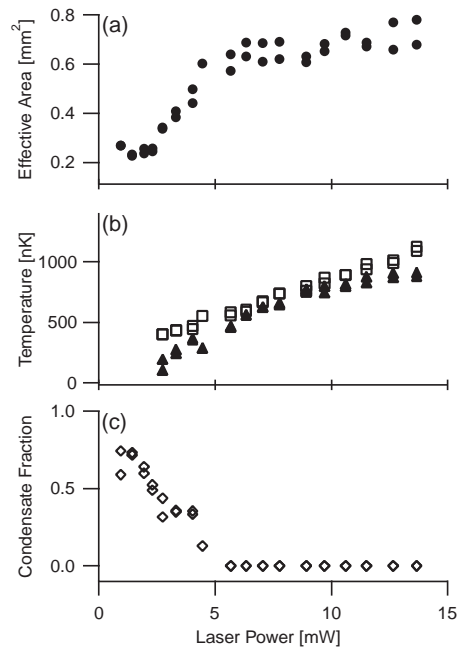}
    \caption[Title]{Characterization of the phase transition.  (a) The effective area is
    the integrated optical density divided by the peak optical density of the absorption
    images. The ``sudden shrinking'' of the area coincides with the appearance of a bimodal
    density distribution and indicates the BEC phase transition. (b) The radial (squares)
    and axial (triangles) temperatures varied smoothly
    during the evaporation. The axial temperatures are in arbitrary
    units. (c) Condensate fraction. Below the critical temperature, the condensate fraction
    increased from zero to up to 75 \%.
    \label{fig:combo}}
    \end{center}
\end{figure}

The phase transition could be identified by plotting the effective area
of the cloud vs.\ laser power $P$ (Fig.~\ref{fig:combo}(a)). At the
phase transition, there was an abrupt change in slope whereas the
temperature changed smoothly.  For a classical gas, the area depends
only on temperature and trap frequencies, which vary smoothly with
$P$.

Quantitative information on temperature, total atom number, and
condensate fraction was obtained by fitting axial profiles (like in
Fig.~\ref{fig:waterfall}) using a bimodal distribution: a
Bose-Einstein distribution for the broad normal component and a
Thomas-Fermi distribution for the narrow (condensate) component.
Condensates containing up to 900,000 molecules and condensate
fractions of up to 75\% were obtained.  The onset of BEC was observed
at a temperature of 600 nK with 1.4 $\times 10^6$ molecules. For an
ideal gas with this number of molecules, the predicted BEC transition
temperature $T_C = 0.94\, \hbar \bar{\omega}\, N^{1/3}/k_B$ is 650
nK,  where $\bar{\omega}$ denotes the geometric mean of the three
trapping frequencies.  This agreement is fortuitous, given the
uncertainty in the trap frequencies at low power \cite{footnote1}.

The cool down is characterized in Fig.~\ref{fig:combo}. Axial
temperatures were obtained from the spatial extent of the thermal
cloud (the size stayed almost constant during the time-of-flight).
The axial temperatures are in arbitrary units \cite{footnote1}.
Absolute radial temperatures were obtained from the ballistic
expansion.  All of our temperature measurements are consistent with
equilibration in three dimensions.

From the expansion of the largest condensates, we determined the mean
field energy $\mu$ to be 300 $\pm$ 100 nK.  The peak density $n$ of
the condensate was obtained from the fit to be $7 \times 10^{13}$
cm$^{-3}$.  The relation $\mu=4 \pi \hbar^2 a\,n/m$, where $m$ is the
molecular mass, implies a molecular scattering length of $a=8$ nm
with an uncertainty of $\pm$ 60 \%.  These uncertainties were estimated from the discrepancy of fits done at different times of flight which were not fully consistent. This might reflect asymmetries and anharmonicities of the trapping potential which were not characterized.

It was predicted that the molecular scattering length $a$ is 0.6
times the atomic scattering length $a_a$ \cite{petr03dimers}. At our
magnetic field of 770$\,$G, the predicted value of $a_a$ lies between
140 and 340 nm depending on the uncertain position of the Feshbach
resonance between 810 and 850 G \cite{gupt03rf}.  The discrepancy
between the predicted and observed values of $a$ needs further study.

The molecular gas decayed faster than extrapolated from
Refs.~\cite{cubi03,joch03lith}.  Just above $T_c$, the thermal cloud
had a peak density of $1 \times 10^{13}$ cm$^{-3}$ and an initial
decay time of about 1 s. Condensate numbers decayed to one third
after a hold time of 30~ms.  Those short lifetimes may reflect
leakage or heating in our optical dipole trap at low laser power. In
the present experiments, the laser power was not stabilized.  The
lifetime of the thermal gas is much longer than estimated values of
the axial trap period of 100 ms and of the collision time of 2 ms,
which should lead to full equilibrium.  Depending on how the
condensate grew during the evaporative cooling, its lifetime may have
been too short to develop phase-coherence in the axial direction
\cite{shva02non_eq_bec}.

Using the experimentally determined scattering length, we find that
the molecular cloud at $T_c$ has a ratio of mean-free path to radius
close to 10 and should show only negligible anisotropy during ballistic
expansion \cite{shva02non_eq_bec}.  Therefore, the onset of strongly
anisotropic expansion is a distinguishing feature of the molecular
condensate (Fig.~\ref{fig:2D_images}).

The \li$_2$ molecules are extremely weakly bound.  The molecular
binding energy depends on the atomic scattering length $a_a$ as
$\hbar^2/m a_a^2$ \cite{kohl03feshbach}.  For an assumed $a_a=
200$\,nm the binding energy is 2 $\mu$K. The molecular state which
causes the Feshbach resonance is the $X\,^1\Sigma^+_g, v=38$ state.
This state is tightly bound, but near the Feshbach resonance it is
strongly mixed with the state of the colliding atoms, and the
molecular wavefunction is spread out over an extension of order
$a_a/2$ \cite{kohl03feshbach}.

Direct evidence for the large size of the molecules was obtained by
resonant imaging during ballistic expansion at high magnetic fields.
At 770 G, molecules could be directly imaged using light in resonance
with the atoms at the same field.  The absorption was comparable to
the zero-field absorption signal obtained after dissociating the
molecules. This shows that the molecular bond is so weak that the
absorption line is shifted from the atomic line by less then a
linewidth $\Gamma$. The molecules are expected to absorb most
strongly near the outer turning point $R$. The excited state
potential is split by the resonant van der Waals interaction $\zeta
\hbar \Gamma (\lambdabar/R)^3$ where $\lambdabar$ is the resonant
wavelength divided by $ 2 \pi$, and $\zeta$ is $\pm 3/4, \pm 3/2$ for
different excited molecular states. The observed absorption signal
implies a molecular size $R \ge$ 100 nm.  It is this huge size
compared with the much smaller size of the molecule in lower
vibrational states which, together with Fermi statistics, inhibits
vibrational relaxation and leads to the long lifetimes
\cite{petr03dimers}.

Condensates were observed after evaporative cooling at various
magnetic fields between 720 and 820 G.  At the lower magnetic fields,
the condensate expanded more slowly, consistent with the predictions
of a smaller repulsive mean-field energy.

In future work we plan to use the molecular BEC as the starting point
for studying the BEC-BCS crossover \cite{becbcs}. By ramping up the
magnetic field across the Feshbach resonance, the molecules are
dissociated into fermionic atoms and the interaction between the
atoms changes from repulsive to attractive allowing for the formation
of Cooper pairs. Starting with an almost pure condensate and
conserving entropy, a Fermi sea should form with temperatures well
within the range where BCS type superfluidity has been predicted
\cite{carr03,mils02}.

In conclusion, we have realized Bose-Einstein condensates of up to
900,000 molecules by evaporative cooling of a spin mixture of
fermionic lithium atoms.

Note added in proof:   In an optical trap with a slightly enlarged
beam waist, we were recently able to hold molecular condensates for
up to 400~ms (or three axial trapping periods) which should result in
3D equilibration. The 1/e decay time was about 200 ms.

This work was supported by the NSF, ONR, ARO, and NASA. We thank A.
Leanhardt for helpful comments.  S.\ Raupach is grateful to the Dr.
J\"urgen Ulderup foundation for a fellowship.


\begin{thebibliography}{10}
\bibitem{footnote2}
This work inferred the presence of a condensate from the number of
particles in a shallow trap, and the magnetic field dependence of the
loss rate consistent with mean field effects. No observation of a
phase transition or measurements of equilibration, temperature or
condensate fraction were reported.
\bibitem{footnote1}
The trap frequencies were scaled from measurements done at full
power. The axial frequency was affected by small magnetic field
curvatures which were negligible at high laser power.
\end{thebibliography}

\end{document}